\PassOptionsToPackage{table}{xcolor}

\documentclass[manuscript]{acmart}
\usepackage{booktabs}   
\usepackage{siunitx}    
\usepackage{tabularx}   
\usepackage{multirow}   

\usepackage{xspace}
\newcommand{\UQU}{\textsc{UnivA}\xspace}
\newcommand{\UmmAlQura}{\textsc{UnivA}\xspace}

\newcommand{\QU}{\textsc{UnivB}\xspace}
\newcommand{\Qassim}{\textsc{UnivB}\xspace}

\newcommand{\KU}{\textsc{UnivC}\xspace}
\newcommand{\Kuwait}{\textsc{UnivC}\xspace}

\newcommand{\HU}{\textsc{UnivD}\xspace}
\newcommand{\Hashemite}{\textsc{UnivD}\xspace}

\AtBeginDocument{%
  \providecommand\BibTeX{{%
    \normalfont B\kern-0.5em{\scshape i\kern-0.25em b}\kern-0.8em\TeX}}}

\setcopyright{acmlicensed}
\copyrightyear{2026}
\acmYear{2026}
\acmDOI{XXXXXXX.XXXXXXX}

\acmConference[Conference acronym 'XX]{Make sure to enter the correct
  conference title from your rights confirmation email}{June 03--05,
  2018}{Woodstock, NY}
 
\acmISBN{978-1-4503-XXXX-X/18/06}

\graphicspath{{./images/}}

\begin{document}

\title{Trust in AI among Middle Eastern CS Students: Investigating Students' Trust and Usage Patterns Across Saudi Arabia, Kuwait and Jordan}

\author{Saleh Alkhamees}
\email{saalkham@cougarnet.uh.edu}
\affiliation{%
  \institution{University of Houston}
  \country{Houston, TX, USA}
}

\author{Ali Alfageeh}
\email{aalfagee@cougarnet.uh.edu}
\affiliation{%
  \institution{University of Houston}
  \country{Houston, TX, USA}
}

\author{Bader Alkhazi}
\email{bader.alkhazi@ku.edu.kw}
\affiliation{%
  \institution{Kuwait University}
  \country{Kuwait}
}

\author{Duaa Alshdaifat}
\email{daalshda@cougarnet.uh.edu}
\affiliation{%
  \institution{University of Houston}
  \country{Houston, TX, USA}
}

\author{Amin Alipour}
\email{maalipou@central.uh.edu}
\affiliation{%
  \institution{University of Houston}
  \country{Houston, TX, USA}
}

\renewcommand{\shortauthors}{Saleh Alkhamees et al.}
\begin{abstract}

\textbf{Background and Context.} Artificial intelligence (AI) tools have been reshaping computing and computer science education. 
Trust in AI is a determining factor in the adoption of these tools.
Recent studies have shown different trust factors across gender and first-generation status among students. However, these studies have focused mainly on Western, Educated, Industrialized, Rich, and Democratic (WEIRD) populations, and their generalizability to other populations with different languages and cultures is unclear. 

\noindent \textbf{Objective.} This study aims to evaluate trust in AI among Middle Eastern computer science students and the factors that can impact it. 

\noindent \textbf{Method.} We replicate a recent study of trust in four universities in three Middle Eastern, Arabic-speaking countries: Saudi Arabia, Kuwait, and Jordan. We analyze trust among students across different factors such as gender and first-generation status.

\noindent \textbf{Findings.} Our results suggest that language fluency can predict trust in AI. Moreover, unlike the results from the US population where female students tended to trust AI more than their male peers, female students in Saudi Arabia indicated lower trust compared to their male counterparts, and we did not observe any noticeable differences across gender in the other countries. We also found a generally negative correlation between English language proficiency and students' confidence. 

\noindent \textbf{Implications.} 
This study highlights differences in students' adoption and trust in AI even within the same region. It emphasizes the need for more investigation into students' adoption and interaction in non-WEIRD regions for equitable adoption of this technology. It also suggests a need for efforts in designing effective AI systems tailored to the cultural and linguistic needs of the region.

\end{abstract}



\maketitle
\section{Introduction}
Generative AI\footnote{Due to outsized impact of generative AI, we use the abbreviation AI instead of generative AI throughout the paper.} has become a mainstream technology that offers a wide range of capabilities including text generation, brainstorming, and programming. 
It affords a wide array of features for computer science education, such as tutoring, debugging, and pair programming. These features can enhance and facilitate students' learning. In the past few years, an extensive body of work has focused on the application and adoption of this technology in CS education.  

Many researchers consider AI models as a promising system for improving learning processing. 
They believe that these models can be used to provide tutoring capabilities, as well as to offer personalized assessment for students. These technologies allow students to enhance their learning \cite{chen_systematic_nodate, garcia-mendez_review_2025, pelaez-sanchez_impact_2024, hartley_artificial_2024, phung_generative_2023}. In this context, Hou et al. \cite{hou_all_2025} found that ChatGPT and other AI tools are seen as the evolutionary education path, providing a variety of benefits, such as intelligent teaching assistance, helping computer science students debug their code and find bugs, and providing personalized explanations. Likewise, Hartley et al. \cite{hartley_artificial_2024} suggest that these tools can tackle the problems of limited resources associated with human tutoring, by providing adaptive learning experiences. Furthermore, in the higher education context, Peláez-Sánchez et al. \cite{pelaez-sanchez_impact_2024} integrate AI with the Education 4.0 framework by encouraging more independent, collaborative, and interactive learning experiences.

Recently, studies show a rapid and significant increase in the adoption of these tools. For instance, Amoozadeh et al. \cite{amoozadeh_trust_2024} found that students generally see AI models as helpful enhancements to their learning process. They highlight the rapidly and significantly increasing adoption of AI tools among students. Their survey at a US university found that most students surveyed had heard of and used AI tools. Other surveys aligned with these findings, such as the one by Dickey et al. \cite{dickey_innovating_2024}, which found that at least 48.5\% of students were using AI tools for homework assistance. This indicates that these tools are easy to access and that students are starting to rely more on them, which may change how they complete their assignments or solve problems over time. Moreover, Raihan et al. \cite{raihan_large_2024} provided a comprehensive systematic review of the literature on AI in CS education and found that more than 80\% of the articles reviewed focused on the undergraduate level. 

However, academic integrity concerns are considered one of the major challenges of using AI, as they bring risks to the educational environment, such as plagiarism, dishonesty in assessments, and the need for updated educational policies to ensure ethical use \cite{tayan_considerations_2024, pelaez-sanchez_impact_2024}. Tayan et al. \cite{tayan_considerations_2024} further investigated this concern by assessing ChatGPT-3.5’s performance on undergraduate computer science course questions covering areas such as programming and cybersecurity, finding an overall accuracy score of 70\%. Nevertheless, incorrect answers were mainly related to questions requiring calculations, detailed answers, and reasoning. The rapid use of AI in education makes it harder to understand whether students are truly grasping the course material or just relying on AI to complete their assignments.

Crucially, Linxen et al. \cite{WEIRD} studied papers published in the CHI proceedings between 2016 and 2020. Their analysis found that 73\% of these articles are based on samples from Western participants, which means that these articles represent less than 12\% of the world’s population. Furthermore, they believed that most of the papers' samples tend to come from industrialized, rich, and democratic countries. Therefore, they motivated us to explore more areas and participants that are not part of the 12\% of the world’s population.

Although some localized studies exist in the Middle East on general AI usage and student performance in some courses in Saudi Arabia and Oman \cite{tayan_considerations_2024, arbab_students_2024}, there is a lack of systematic investigation into the extent of AI usage, its overall perception, and the educational factors that affect CS students within the broader Middle Eastern region. General existing studies serve as crucial perspectives but do not focus on the nuances of courses in programming education or the specific impact on computer science students. Therefore, findings from Western countries cannot be directly applied to the Middle East. Different cultural views on authority, local education policies, and language barriers make the situation unique which requires specific investigation and this research aims to address this gap.

In this paper, we aim to address this gap and provide more details on this issue. We replicated a survey among different students from three Middle Eastern countries. We targeted two Saudi Arabian universities and one university from Kuwait and Jordan. The purpose of this study is to identify significant differences between students from these three countries. In addressing our three research questions, this study makes the following key contributions:
\begin{enumerate}
    \item It measures how Middle Eastern students use AI, with almost all having tried it (98\%), while maintaining a healthy level of trust, using AI as a helper rather than an expert.

    \item It shows how the educational environment affects trust, showing that the gender gap in AI trust is not a universal cultural attribute of the region but is correlated with gender segregated environments in Saudi Arabia versus mixed gender environments in Kuwait and Jordan.
\end{enumerate}

The following three research questions guided our study:
\begin{itemize}

\item \textbf{RQ1:} How widely do Middle Eastern students use AI, and what do they see as its advantages and disadvantages?

\item \textbf{RQ2:} To what extent do Middle Eastern students trust AI tools?

\item \textbf{RQ3:} How do Middle Eastern students perceive AI's impact on their programming learning?
\end{itemize}

In addressing these three research questions, our study aims to explore how Middle Eastern students interact with AI tools in programming education. We investigate the prevalence of usage and perceived advantages or disadvantages, as well as the level of trust students have in these tools. Moreover, we aim to explore students' perceptions of AI's impact on their learning. The goal of this study is to contribute to computing education research, especially in the Middle Eastern region, by highlighting the factors that influence students’ adoption and effective use of AI tools.

\section{ Related Work}

\subsection{AI in Education}

Some studies suggest students often use AI as a digital peer \cite{phung_generative_2023} or as a personalized tutor \cite{prather_widening_2024, bassner_iris_2024}. Beyond basic assistance, other researchers use AI to support activities such as automatic question generation \cite{garcia-mendez_review_2025} and providing in-depth explanations \cite{dickey_innovating_2024}. 

Bassner et al. \cite{bassner_iris_2024} went a step further and introduced Iris, a pedagogical assistant used in a large programming course. It is designed to act as a tutor in a way that is pedagogically effective by guiding students without revealing complete answers. However, from a research perspective, although these tools offer some educational benefits, their impacts still depend on how students engage with the learning process rather than relying solely on the guidance generated by AI tools.

In the Middle East, limited data related to programming exists. For example, Arbab et al. \cite{arbab_students_2024} conducted a study in Oman and found that over 50\% of surveyed students considered the use of AI tools to be improperly regulated by institutions. This lack of clarity led to diverse student opinions; some wanted more space for using newly developed technology, while others desired stricter policies. Furthermore, penalties imposed by institutions included grade reduction (35\%), resubmission (30\%), warnings (26\%), and failing grades (10\%). Similarly, a study in Saudi Arabia aimed to obtain recommendations for regulating AI tools in higher education, especially in technology courses \cite{tayan_considerations_2024}. Although these studies have contributed valuable insights, we still do not fully understand how students use and trust these tools in Middle Eastern countries. 

Beyond these localized Middle Eastern studies, Haque and Hundhausen \cite{Haque} conducted a comparative study examining AI adoption across two different regions, revealing significant disparities in access and usage patterns. Their survey targeted 534 undergraduate computing students across six universities: 250 students in the United States across three universities, and 284 students in Bangladesh across three universities. Their findings revealed substantial differences in AI access, usage behavior, and student perceptions. 95\% of Bangladeshi students reported using AI assistants for academic purposes compared to 64\% of US students, even though Bangladeshi students faced considerable barriers, including cost limitations. For instance, 51.6\% of them mentioned this challenge compared to 8.6\% of US students. Additionally, they cited inconsistent internet connectivity and limited knowledge of how to access reliable AI tools. Interestingly, only 5.1\% of Bangladeshi students paid for AI subscriptions versus nearly 25\% of US students. The study revealed that the digital divide has a significant impact on how students in different socioeconomic contexts engage with AI tools. It highlights that findings from Western contexts cannot be applied directly to other regions without considering local infrastructural, economic, and educational factors.

In this paper, we aim to obtain more detailed insights, which will help us provide more reliable conclusions about students in the Middle Eastern region.

\subsection{AI in Programming Education}

AI models are heavily adopted in introductory programming courses (CS1) \cite{raihan_large_2024}. Raihan et al. \cite{raihan_large_2024} noted that more than half of their reviewed studies (65 studies) focused on introductory programming; these tools are widely utilized to generate and implement code, identify and resolve errors, generate practice questions, and provide formative feedback (e.g., \cite{sarsa_automatic_2022, amoozadeh_student-ai_2024, jacobs_evaluating_2024, raihan_large_2024, cambaz_use_2024}). Students appreciate these models for acting as effective virtual tutors, providing explanations that are easy to understand \cite{raihan_large_2024}. 

Notably, Amoozadeh et al. \cite{amoozadeh_trust_2024} found that students mostly use AI to seek programming help and to understand existing code instead of writing new code entirely. However, Alpizar-Chacon et al. \cite{alpizar-chacon_students_2025} reported that students use AI heavily in their projects, primarily to generate code and ideas in an advanced web development course. Dickey et al. \cite{dickey_innovating_2024} conducted a survey supporting these findings of high engagement and discovered that at least 48.5\% of students use AI for homework.

However, the efficacy of AI tools remains mixed. Some studies demonstrate that AI can solve basic problems, repair bugs, and write better code. Yet, other benchmarks comparing AI tools to human tutors revealed that even state-of-the-art models struggle in certain settings. Generative AI systems often produce code containing minor errors that require human expertise to detect and fix. Thus, educators are encouraged to help students understand the limitations of AI tools. They should demonstrate how AI model outputs often contain minor errors that, while easily managed by instructors, pose risks for novice learners \cite{dickey_innovating_2024, cambaz_use_2024}.

Recent research has highlighted concerning impacts on learning outcomes. A study by Jost et al. \cite{jost_impact_2024} emphasizes one of the critical concerns regarding AI use: a significant negative correlation between increased reliance on AI models and lower final grades in terms of critical thinking. This occurred especially when students used AI for intensive tasks, such as code generation and debugging. Furthermore, Prather et al. \cite{prather_widening_2024} observed that novice programmers often struggle with metacognitive difficulties when using AI, leading to an "unwarranted illusion of competence." Heavy reliance on AI tools can potentially exacerbate the gap between high-performing students who use these tools to accelerate their work and struggling students who are hindered by them. As a result, students may overestimate their understanding, potentially hindering the development of critical programming skills. This creates a dangerous cycle: novice students may lack the experience to verify AI-generated code but trust it because it seems correct, leading to the adoption of poor coding habits that are difficult to correct later.

For Middle Eastern students, programming education research intersects significantly with linguistic challenges. A study by Prather et al. \cite{prather_breaking_2024} conducted with fluent speakers of Arabic, Chinese, and Portuguese found that although multilingual prompting can empower non-native English speakers, Arabic speakers achieved the lowest success rate (27\%) in generating code. This outcome, likely linked to the limited availability of Arabic training data, highlights the critical need for better localized AI tools. Consequently, Middle Eastern students face two main challenges: understanding difficult programming concepts and dealing with AI tools that perform poorly in their native language, often necessitating a switch to English and increasing their cognitive load.

Most importantly, existing research on programming education regarding AI usage focuses primarily on Western, Educated, Industrialized, Rich, and Democratic (WEIRD) dimensions \cite{WEIRD}. While some research shows that Arabic speakers experience profound difficulty with AI for code generation \cite{prather_breaking_2024, pirzado_evaluating_2025}, there is a gap in integrated research quantifying the extent of AI usage and the explicit perceptions of Middle Eastern CS students regarding efficiency, learning outcomes, and ethical concerns. For example, language limitations might lead students to copy English solutions without full understanding, or deter them from using AI entirely.

\subsection{Trust in AI Tools}

 Trust was defined as "the user’s judgment or expectations about how the AI system can help when the user is in a situation of uncertainty or vulnerability" by Amoozadeh et al. \cite{amoozadeh_trust_2024}. Trust influences the adoption and appropriate use of AI: where undertrust can lead to abandonment of useful help, while overtrust can lead to uncritical acceptance for any AI’s suggestions. In the context of computer science education, this is a critical issue. Trusting AI too much is risky because it often produces correct-looking code that contains logical errors, causing students to focus on making the code work rather than learning the underlying logic.

To understand how this trust forms, Pitts et al. \cite{pitts_understanding_nodate} suggested that students often develop a distinct form of human-AI trust that differs from human-technology models or even with the traditional human-human. If people trust the individuals behind the system, they are more likely to use it (trust in people). Additionally, if people trust the technology itself, they are more likely to believe it is useful and plan to use it (trust in technology). Amoozadeh et al. \cite{amoozadeh_trust_2024} targeted students in the USA and India. Their results showed that students have varying levels of trust in AI, with approximately 47\% reporting trust, 36\% remaining neutral and 16\% expressing distrust. The result indicated that there is a moderate positive correlation between trust in AI and students' feelings of confidence and motivation when using these tools. Existing studies indicate that trust in AI tools varies among students from different cultural backgrounds. While students from Western countries may easily trust AI outputs, students from more conservative regions, like Middle Eastern countries, tend to use these tools with more caution.

Amoozadeh et al. \cite{amoozadeh_trust_2024} claimed that even students who believe that AI is helpful to them may still express distrust and a need for human oversight. Notably, this correlation between trust and improving motivation is stronger among first-generation students compared to continuing-generation students. The survey revealed that US students were less skeptical about AI output being as good as a highly competent person compared to Indian students. This difference means that trust in AI is not universal; it depends on culture. Thus, if culture affects how students perceive AI abilities, then using Western ideas about trust with Middle Eastern students, who have different educational values, may lead to wrong conclusions.

Moreover, Demidova et al. \cite{demidova_john_2024} found that research has also uncovered subtle biases where AI models tend to present winners as those speaking the primary language the model is prompted with. Because of this, trustworthiness must be evaluated in different cultural contexts outside of the Western world. Addressing this need, a benchmark called "AraTrust" was introduced to evaluate the trustworthiness of Arabic LLMs across different dimensions such as safety, privacy, and ethics, revealing significant performance differences between open-source and proprietary models. Because of this language bias, Arab students might start to think that good technical solutions only come in English, which may affect how much they trust AI when using their native language.

While existing research on student trust has primarily focused on students in the US and other developed countries \cite{amoozadeh_trust_2024, pitts_understanding_nodate}, there is a lack of empirical research specifically investigating other regions of the world. Therefore, a comprehensive study is needed to investigate more on trust and its relationship with motivational factors among computer science students in the Middle East, considering their cultural and educational systems. Such research is essential to understand and shape the trust levels and usage behaviors among them.

\subsection{Cultural and Linguistic Aspects of AI models}
The performance gap is evident. Pirzado et al. \cite{pirzado_evaluating_2025} discovered that AI models generally perform better with English inputs. For non-native English speakers, learning programming is challenging because almost all programming languages and documentation are in English, which creates a high cognitive barrier. Studies show that AI models struggle when processing non-English prompts, especially when it comes to programming tasks. Arabic students seem to face greater challenges, achieving lower accuracy with only a 27\% success rate and experiencing more model misinterpretations \cite{prather_breaking_2024}. The reason for this is likely the limited availability of Arabic datasets for LLM training. AI models tend to perform better with English prompts, even when used by non-English speakers \cite{prather_breaking_2024, pirzado_evaluating_2025}. As a result, this may drive Arabic students to use English, despite their limitations in the language, which can lead to further challenges. 

Other studies have focused on AI models as potential compilers for Arabic programming \cite{sibaee_llms_2024} and evaluated AI models for Arabic code summarization \cite{aljohani_evaluating_2025}. These efforts aim to bridge the gap for Arabic-speaking students, who face challenges while using AI models in their native language. Moreover, Sukiennik et al. \cite{sukiennik_evaluation_2025} revealed that AI models exhibit a Western cultural bias, mostly due to the sources of the datasets used in training. Quantifying this bias, their study shows that LLM models, such as GPT-3 and GPT-4, tend to display Western values. Expanding on this critical issue, Schiff et al. \cite{schiff_education_2022} further argued that this emphasizes the critical need for culturally sensitive systems that account for local vocabulary and customs, particularly when implementing AI in tutoring systems. This is described as a potential "portability trap" when systems trained on data from high-income Western institutions are implemented in different sociotechnical contexts. They argue that this area needs more studies to evaluate cultural value alignment systematically across models, countries, and languages. The literature acknowledges that the gap in accessing technology—the "digital divide"—remains a significant obstacle, as paid AI models may limit usage for entire populations, which may particularly affect large portions of the global population \cite{pelaez-sanchez_impact_2024, cambaz_use_2024}.

Despite evidence demonstrating the precise difficulties in Arabic language performance, cultural bias, and regional policy gaps (e.g., \cite{prather_breaking_2024, demidova_john_2024, schiff_education_2022}), the impact of generating code tools on non-English programming education remains unexplored. 
In this context, no large-scale empirical research has fully investigated the connection between these confirmed linguistic/cultural performance factors and the resulting student use, trust, and perceptions of AI systems among Middle Eastern computer science students. Thus, there is a need for more research addressing Arabic language usage in programming courses to help bridge the knowledge gap and improve AI capabilities in this area. Crucially, it still remains unknown whether these factors cause students in the Middle East to entirely abandon AI tools or adopt them with misplaced trust that ignores their insufficiency. This study focuses on this specific demographic area, which can help bridge that gap by providing localized data on how students recognize these factors and translate them into real-world behavior, trust adjustment, and ultimate academic outcomes.

\section{Methodology}

\subsection{Data Collection}
To better understand students’ perspectives, we conducted the survey in CS classes, instead of distributing it online or using students' university emails. We believe that students can engage more by doing it this way and provide more reliable responses. The survey was conducted during the Fall 2025 semester across a variety of CS courses. Since all the students' native language is Arabic, we conducted the survey in Arabic and then translated participants' feedback into English for further analysis, which will be discussed later in this paper. Conducting the study across three countries allowed us to collect more data, which helped us make more generalizable conclusions. Also, we targeted students from Middle Eastern countries and Arabic-language speakers to gain a better understanding of the end-users there, compared to the Western, Educated, Industrialized, Rich, and Democratic (WEIRD)\cite{WEIRD}.

\subsection{Survey Design}
In this study, we follow the study design of Amoozadeh et al. \cite{amoozadeh_trust_2024}, which was adapted from Körber's Trust in Automation survey \cite{korber2019}. 
This study uses a survey to understand trust in AI among CS students. We adopted the same survey translated into Arabic. Since Arabic has several dialects, we validated the clarity and validity of the survey through a pilot study with 17 Arabic-speaking participants with diverse backgrounds to ensure the questions were understandable to students. This data was not included in our analysis.

\subsubsection{Survey Questions}

Our survey includes a variety of open and closed-ended questions, starting with demographic information, followed by the participants' confidence in programming, their trust in AI, and their usage and opinions about AI tools. The demographic questions contained a question about the experience of participants using AI in programming. We aimed to classify students into three categories:
\begin{itemize}

\item First Category: for students who had used AI tools. Their experience and feedback were collected as well as their trust in the tools. Participants' motivation and confidence in programming were evaluated.

\item Second Category: for students who had heard about AI tools, but had never used them. We asked students if they were willing or interested in trying AI tools and their trust in them.

\item Third Category: for students who had never heard about AI tools. Here, the participants had no experience using AI at all. We asked them general questions including their trust in AI tools and their awareness and impressions of these technologies.
\end{itemize}

The last section targeted all participants in the survey. We asked them about their opinions on using AI, especially in programming. Our goal with the diversity of questions was to collect comprehensive data from students regarding their experience with AI tools. Most of the questions in the survey were multiple-choice, so our analysis was mainly quantitative. For the open-ended question, we leveraged the open-coding structure \cite{Corbin} to determine and condense themes in the students’ responses.

\subsubsection{Participants}
The survey was distributed to undergraduate and graduate students at \UmmAlQura  and \Qassim in Saudi Arabia, \Kuwait in Kuwait and \Hashemite in Jordan. They are public universities in the corresponding countries.
These countries are Arabic-speaking countries.
We selected these countries because they share the same language and cultural background. 
This helped us to obtain diverse perspectives within a common linguistic and cultural framework.

\subsubsection{Course Selection}
Our target was to distribute the survey among CS students. We started with \UQU students by conducting the survey in two different classes: Computer Programming 1 and Computer Programming 2. Then we conducted it at \QU in three different classes: Computer Programming 1, Computer Programming 2 and Data Structures. At \UQU and \QU, the classes are typically separated by gender, with male and female students attending different classrooms. In contrast to that, at \HU, we distributed the survey in three different classes: Object-Oriented Programming, Machine Learning, and Artificial Intelligence. At \KU, we distributed the survey in three different classes: Information Technology, Database, and Programming and Problem Solving.

\subsection{Data Analysis}
\subsubsection{Quantitative Analysis}
Our aim is to understand students' trust in using AI. Therefore, after collecting the data, we utilized descriptive statistics, correlations, and comparisons of distributions to obtain various factors related to our objectives, which are trust and usage.

\subsubsection{Qualitative Analysis}
Our aim is to gain more details about students' perceptions in general about AI tools. At the end of the survey, there was an optional open-ended question for the students. The question was: \textbf{\textit{In general, how do you feel about AI systems in programming?}}. After collecting the data, we analyzed each student's response using thematic open coding. We provide more details in the Results section.

\section{Results}
\subsection{RQ1: Exposure, Adoption, and Usage of AI}
\subsubsection{Participant Demographics}
A total of 319 survey responses were initially collected from four locations. In Saudi Arabia, \UQU contributed 54 responses and \QU contributed 73 responses. In Jordan, \HU contributed 130 responses. In Kuwait, \KU contributed 62 responses. During the data verification process, responses were reviewed for completeness and validity. We found some were invalid or incomplete. The number of such responses at each university was as follows: \UQU (rejected = 1, incomplete = 18), \QU (rejected = 2, incomplete = 26), \HU (rejected = 2, incomplete = 50), and \KU (rejected = 2, incomplete = 16). As a result, a total of 7 responses were rejected and 110 were identified as incomplete across all four locations. Our final dataset includes 202 valid responses, representing an overall retention rate of approximately 63\%. Table \ref{tab:Demographics} presents the participant demographics of these 202 students across three Middle Eastern countries and four universities.

\begin{table}[htbp]
  \caption{Demographic characteristics of participants ($N=202$)}
  \label{tab:Demographics}
  \centering
  \footnotesize
  \renewcommand{\arraystretch}{1.2} 
  \setlength{\tabcolsep}{6pt} 
  
  \begin{tabular}{llcc}
    \toprule
    \textbf{Characteristic} & \textbf{Category} & \textbf{\textit{n}} & \textbf{\%} \\
    \midrule
    
    \rowcolor{gray!10}
      & \textbf{Saudi Arabia} & \textbf{80} & \textbf{39.6} \\
      & \hspace{3mm} Male & 36 & 17.8 \\
    \rowcolor{gray!10}
      & \hspace{3mm} Female & 44 & 21.8 \\
      & \textbf{Jordan} & \textbf{78} & \textbf{38.6} \\
    \rowcolor{gray!10}
      & \hspace{3mm} Male & 42 & 20.8 \\
      & \hspace{3mm} Female & 36 & 17.8 \\
    \rowcolor{gray!10}
      & \textbf{Kuwait} & \textbf{44} & \textbf{21.8} \\
      & \hspace{3mm} Male & 19 & 9.4 \\
    \rowcolor{gray!10}
    \multirow{-9}{*}{\textbf{Country}} 
      & \hspace{3mm} Female & 25 & 12.4 \\
    \midrule
    
      & \Qassim University (Saudi) & 45 & 22.3 \\
    \rowcolor{gray!10}
      & \UmmAlQura University (Saudi) & 35 & 17.3 \\
      & \Hashemite University (Jordan) & 78 & 38.6 \\
    \rowcolor{gray!10}
    \multirow{-4}{*}{\textbf{University}} 
      & \Kuwait University (Kuwait) & 44 & 21.8 \\
    \midrule
    
      & Female & 105 & 52.0 \\
    \rowcolor{gray!10}
    \multirow{-2}{*}{\textbf{Gender (Total)}} 
      & Male & 97 & 48.0 \\
    \midrule
    
      & Continuing-generation & 138 & 68.3 \\
    \rowcolor{gray!10}
    \multirow{-2}{*}{\textbf{\shortstack[l]{First-Generation\\Status}}} 
      & First-generation & 64 & 31.7 \\
    \midrule
    
      & UG 3rd year & 90 & 44.6 \\
    \rowcolor{gray!10}
      & UG 2nd year & 33 & 16.3 \\
      & UG 4th year & 31 & 15.3 \\
    \rowcolor{gray!10}
      & Post-graduate & 18 & 8.9 \\
      & UG 5th year+ & 18 & 8.9 \\
    \rowcolor{gray!10}
    \multirow{-6}{*}{\textbf{Academic Level}} 
      & UG 1st year & 7 & 3.5 \\
    \midrule
    
      & $<$ 1 year & 62 & 30.7 \\
    \rowcolor{gray!10}
      & 1 year & 46 & 22.8 \\
      & 2 years & 47 & 23.3 \\
    \rowcolor{gray!10}
      & 3 years & 24 & 11.9 \\
    \multirow{-5}{*}{\textbf{\shortstack[l]{Programming\\Experience}}} 
      & $\ge$ 4 years & 19 & 9.4 \\
      
    \bottomrule
  \end{tabular}
\end{table}

\subsubsection{Adoption Rates and AI Tool Usage}
Table \ref{tab:Usage} shows, Middle Eastern students tend to adopt AI tools at a very high rate. Of the 202 participants, 198 students (98.0\%) reported having used at least one AI tool, while only 4 students (2.0\%) reported that they knew about them but never use them. No participant reported being completely unaware of AI tools. Among the active users (n=198), it is clear that the vast majority of them (92.9\%) use OpenAI’s ChatGPT. Interestingly, the Chinese AI tool, DeepSeek follows, with a substantial adoption rate of 62.1\%, followed by GitHub Copilot (41.4\%) and Grok (26.3\%). 

In terms of AI tool task usage, the most common tasks were 'Understanding/Explaining code' (84.3\%) and 'Learning new concepts' (77.3\%) followed by 'Modifying existing code' which was reported by 69.7\% of students. This habit indicates that students frequently use AI to improve their own work, rather than totally relying on AI tools for 'writing new code from scratch' (60.1\%) or 'seeking help from other' (36.9\%).

\begin{table}[htbp]
  \caption{AI usage and Task type ($N=198$)}
  \label{tab:Usage}
  \centering
  \small 
  \renewcommand{\arraystretch}{1.3} 
  
  \begin{tabular}{l c c}
    \toprule
    \textbf{Category / Item} & \textbf{Count (\textit{n})} & \textbf{Percentage (\%)} \\
    \midrule
    
    \multicolumn{3}{l}{\textbf{AI Tools Used}} \\
    
    \rowcolor{gray!10}
    \hspace{3mm} OpenAI's ChatGPT & 184 & 92.9 \\
    \hspace{3mm} DeepSeek & 123 & 62.1 \\
    \rowcolor{gray!10}
    \hspace{3mm} Copilot & 82 & 41.4 \\
    \hspace{3mm} GitHub (General/Search) & 57 & 28.8 \\
    \rowcolor{gray!10}
    \hspace{3mm} Grok & 52 & 26.3 \\
    \hspace{3mm} Other & 5 & 2.5 \\
    
    \addlinespace
    
    \multicolumn{3}{l}{\textbf{Tasks}} \\
    
    \rowcolor{gray!10}
    \hspace{3mm} To understand / explain code & 167 & 84.3 \\
    \hspace{3mm} To learn new CS concepts & 153 & 77.3 \\
    \rowcolor{gray!10}
    \hspace{3mm} To modify existing code & 138 & 69.7 \\
    \hspace{3mm} To write new code & 119 & 60.1 \\
    \rowcolor{gray!10}
    \hspace{3mm} To seek programming help & 73 & 36.9 \\
    \hspace{3mm} Writing essays & 42 & 21.2 \\
    
    \bottomrule
  \end{tabular}
\end{table}

\subsubsection{confidence in using AI Tools}
\begin{figure*}[t]
    \centering
    \includegraphics[width=0.9\textwidth]{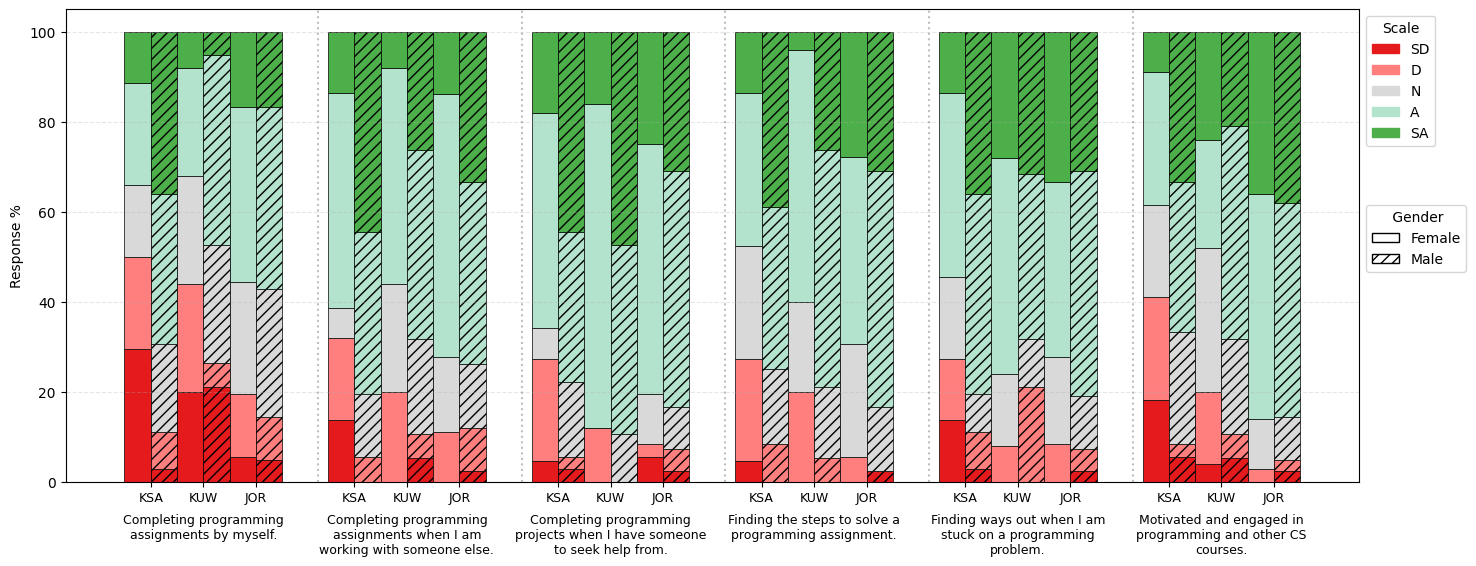}
    \caption{Confidence by gender ($N=202$)}
    \label{fig:CoGender}
\end{figure*}

Responses were measured using the five-point Likert scale and analyzed using the Wilcoxon rank-sum test. As shown in Figure \ref{fig:CoGender}, in Saudi Arabia, male students report feeling more confident than female students across all metrics. In the Kingdom, 25 male students (69.4\%) agreed they are confident in "completing programming assignments by myself," compared to only 15 female students (34.1\%) ($p=0.0001$). In fact, 22 female students (50.0\%) reported disagreeing, indicating low confidence when working alone. Furthermore, 24 male students (66.7\%) agreed they feel more "motivated and engaged in programming" than female students, whereas only 17 students (38.6\%) said the same ($p=0.0008$). It also seems that males feel more comfortable "working with someone else," with 29 students (80.6\%) preferring to work with others, whereas 27 female students (61.4\%) said the same ($p=0.0012$). The majority of male students (75.0\%) feel confident "finding the steps to solve a programming assignment," compared to 21 female students (47.7\%) ($p=0.0023$). Additionally, 80.6\% of male students agreed they can "find ways out when stuck," while only 54.6\% of female students agreed ($p=0.0041$). In Kuwait, females show similar trends to Saudi female students across all metrics. For example, only 60.0\% of of females agreed they can find the steps to solve an assignment, compared to 79.0\% of male students ($p=0.0383$). An interesting result emerged from Jordan: 90.0\% of first-generation students agreed they can "find ways out when stuck on a programming problem," compared to 72.4\% of non-first-generation students ($p=0.0362$). This indicates that first-generation students may have developed greater self-reliance in solving programming problems compared to their peers from more traditional academic backgrounds.

\subsubsection{Experience of Using AI Tools}

\begin{figure*}[t]
    \centering
    \includegraphics[width=0.9\textwidth]{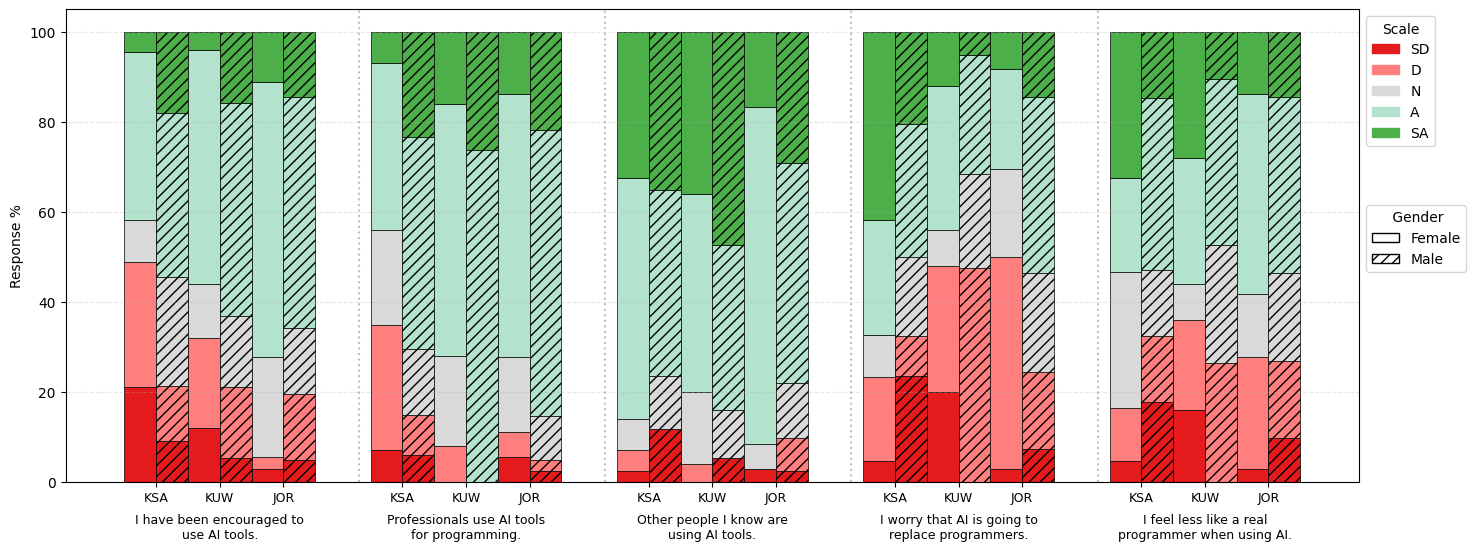}
    \caption{Experience by gender ($N=202$)}
    \label{fig:ExGender}
\end{figure*}

In Figure \ref{fig:ExGender}, among CS students in Saudi Arabia, male students trust AI more than female students in general. 24 male students (70.6\%) agree with the statement that says, "Professionals use AI tools for programming", compared to only 19 female students (44.2\%). This suggests that male students are more likely to perceive AI as a more standard, accepted professional tool ($p=0.0099$). Men tend to feel more encouraged to use AI. Also, 18 male students (54.5\%) agree they have been encouraged to use AI, compared to 21 female students (48.8\%) who disagreed, indicating they feel less encouraged, which indicates that men gained more social support ($p=0.0338$). Again, males worry less about losing jobs. 29 female students (67.4\%) agreed with "I worry that AI is going to replace programmers," compared to 17 male students (50.0\%) ($p=0.0346$). In contrast to Saudi Arabia, 22 Jordanian male students (53.7\%) worry about being replaced by AI, compared to only 11 female students (30.6\%) share that the same worry. This implies that the trust or fear regarding AI depends on the country ($p=0.0399$). In Kuwait, interestingly, 100\% of male students agreed that professionals use AI tools, compared to 72\% of female students ($p=0.0433$). In terms of first-generation students, 44.5\% of first-generation students agreed with the statement "I feel less like a real programmer when using AI", compared to 57.7\% of non-first-generation students ($p=0.0996$). This indicates that students from more academic backgrounds hold stricter traditional views on what "real programming" looks like, making them trust their own skills less when using AI tools for coding.

\subsection{RQ2: Trust in AI Tools}
\subsubsection{Overall Trust Levels}
Six of trust statements adapted from Körber's Trust in Automation scale were rated on a 5-point Likert scale (1=Strongly Disagree, 5=Strongly Agree). Table \ref{tab:Trust} presents the distribution of student responses across all trust dimensions.

Students show moderate levels of overall trust. The average level of trust was 3.11 (SD=0.80) across all six dimensions. Statement S3 had the highest agreement with a mean of 3.46, suggesting that students feel they understand AI behavior patterns. The lowest agreement was for statement S4, with a mean of 2.72, indicating that students do not blindly trust AI outputs without verification, which reflects healthy skepticism. Notably, 98 out of 198 participants (49.5\%) disagreed or strongly disagreed with statement S4. This suggests that students demonstrate critical thinking in their AI usage. Furthermore, 53.5\% of students agreed with statement S5, suggesting that despite moderate trust levels, students find value in these AI tools.

\subsubsection{Trust Variations Across Countries} If we look at each country in Table \ref{tab:Trust}, students from Jordan showed the highest overall trust (Mean=3.24, SD=0.67), followed by Kuwaiti students (Mean=3.20, SD=0.71), and Saudi students (Mean=2.93, SD=0.93). Across all three countries, statement S3 received the highest ratings, while statement S4 received the lowest, maintaining the pattern observed in the student overall.

\begin{table*}[htbp]
    \caption{Distribution of Trust in AI Among Participants ($N=198$)}
    \label{tab:Trust}
    \small 
    \renewcommand{\arraystretch}{1.3} 
    \setlength{\tabcolsep}{2pt} 
    
    \newcolumntype{Q}{>{\raggedright\arraybackslash\hsize=2.8\hsize}X}
    \newcolumntype{L}{>{\centering\arraybackslash\hsize=0.9\hsize}X}
    \newcolumntype{S}{>{\centering\arraybackslash\hsize=0.45\hsize}X}
    \newcolumntype{C}{>{\centering\arraybackslash\hsize=0.6\hsize}X}

    \begin{tabularx}{\textwidth}{ Q L L L L L S S C C C }
        \toprule
        \textbf{Trust Statement} & 
        \textbf{\shortstack{Strongly\\Disagree}} & 
        \textbf{Disagree} & 
        \textbf{Neutral} & 
        \textbf{Agree} & 
        \textbf{\shortstack{Strongly\\Agree}} & 
        \textbf{Mean} & 
        \textbf{\textit{SD}} &
        \textbf{\shortstack{Saudi\\(n=77)}} & 
        \textbf{\shortstack{Kuwait\\(n=44)}} & 
        \textbf{\shortstack{Jordan\\(n=77)}} \\
        \midrule
        
        \rowcolor{gray!10}
        S1: I trust the system's output & 
        24 (12.1\%) & 60 (30.3\%) & 57 (28.8\%) & 47 (23.7\%) & 10 (5.1\%) & 
        2.79 & 1.09 & 2.68 & 2.82 & 2.90 \\
        
        S2: Output as good as competent person & 
        23 (11.6\%) & 49 (24.7\%) & 35 (17.7\%) & 64 (32.3\%) & 27 (13.6\%) & 
        3.12 & 1.26 & 2.94 & 3.23 & 3.23 \\
        
        \rowcolor{gray!10}
        S3: Know what will happen next time & 
        12 (6.1\%) & 20 (10.1\%) & 45 (22.7\%) & 106 (53.5\%) & 15 (7.6\%) & 
        3.46 & 0.99 & 3.26 & 3.61 & 3.58 \\
        
        S4: Believe output when uncertain & 
        21 (10.6\%) & 77 (38.9\%) & 46 (23.2\%) & 45 (22.7\%) & 9 (4.5\%) & 
        2.72 & 1.07 & 2.57 & 2.68 & 2.88 \\
        
        \rowcolor{gray!10}
        S5: Personal preference for using AI & 
        5 (2.5\%) & 30 (15.2\%) & 57 (28.8\%) & 90 (45.5\%) & 16 (8.1\%) & 
        3.41 & 0.93 & 3.21 & 3.66 & 3.48 \\
        
        S6: Overall, I trust the AI system & 
        16 (8.1\%) & 35 (17.7\%) & 63 (31.8\%) & 72 (36.4\%) & 12 (6.1\%) & 
        3.15 & 1.04 & 2.91 & 3.18 & 3.36 \\
        
        \midrule
        \textbf{Overall Trust Score} & \multicolumn{5}{c}{--} & \multicolumn{2}{c}{--} & \textbf{2.93} & \textbf{3.20} & \textbf{3.24} \\
        
        \bottomrule
        \multicolumn{11}{l}{\footnotesize \textit{Note:} \textit{SD} in the eighth column denotes Standard Deviation.}
    \end{tabularx}
\end{table*}

\subsubsection{Trust Differences Across Countries by Gender } If we explore this in more depth, the noticeable gender differences in trust appeared in Saudi Arabia. Unlike the study conducted in the United States by Amoozadeh et al. \cite{amoozadeh_trust_2024}, as shown in Table \ref{tab:trust_gender_country}, Saudi female students reported significantly lower levels of trust than Saudi male students: statement S1 (Female=2.12 vs. Male=3.38, $p < 0.001$), statement S2 (Female=2.47 vs. Male=3.53, $p < 0.001$), statement S3 (Female=2.88 vs. Male=3.74, $p = 0.002$), statement S5 (Female=2.88 vs. Male=3.62, $p = 0.001$), and statement S6 (Female=2.51 vs. Male=3.41, $p < 0.001$). This special pattern appears only in Saudi Arabia, likely due to the separation of the educational environment by gender, which directly influences how male and female students perceive and trust AI technologies. Interestingly, Kuwaiti female students tend to trust AI less than males, similar to females in Saudi Arabia, but without a significant $p$-value. Jordanian females show almost the same level of trust as male students, also with no significant $p$-value.

In terms of first- and continuing-generation students, trust levels between them were largely similar across most dimensions. However, an exception appeared in Jordan. First-generation students reported significantly higher trust in system output: statement S1 (Mean=3.30) compared to continuing-generation students (Mean=2.75, $p=0.043$). This pattern was not observed in Saudi Arabia or Kuwait.

\begin{table}[htbp]
  \caption{Trust Differences Across Countries by Gender}
  \label{tab:trust_gender_country}
  \centering
  \footnotesize
  \renewcommand{\arraystretch}{1.3} 
  \setlength{\tabcolsep}{4pt} 
  
  \begin{tabular}{l l c c c}
    \toprule
    \textbf{Country} & \textbf{Dimension} & \textbf{\shortstack{Female\\Mean (\textit{n})}} & \textbf{\shortstack{Male\\Mean (\textit{n})}} & \textbf{\textit{p}} \\
    \midrule
    
    \rowcolor{gray!10} & S1: Trust output & 2.12 (43) & 3.38 (34) & $<.001$ \\
    & S2: Expert-level & 2.47 (43) & 3.53 (34) & $<.001$ \\
    \rowcolor{gray!10} & S3: Predictability & 2.88 (43) & 3.74 (34) & .002 \\
    & S4: Uncertain & 2.37 (43) & 2.82 (34) & .067 \\
    \rowcolor{gray!10} & S5: Preference & 2.88 (43) & 3.62 (34) & .001 \\
    & S6: Overall trust & 2.51 (43) & 3.41 (34) & $<.001$ \\
    \rowcolor{gray!10} \multirow{-7}{*}{\rotatebox{90}{\textbf{Saudi}}} & \textbf{Overall Index} & \textbf{2.54} & \textbf{3.42} & \textbf{$<.001$} \\
    \midrule
    
    & S1: Trust output & 2.56 (25) & 3.16 (19) & .054 \\
    \rowcolor{gray!10} & S2: Expert-level & 3.16 (25) & 3.32 (19) & .751 \\
    & S3: Predictability & 3.68 (25) & 3.53 (19) & .918 \\
    \rowcolor{gray!10} & S4: Uncertain & 2.44 (25) & 3.00 (19) & .124 \\
    & S5: Preference & 3.44 (25) & 3.95 (19) & .054 \\
    \rowcolor{gray!10} & S6: Overall trust & 3.12 (25) & 3.26 (19) & .892 \\
    \multirow{-7}{*}{\rotatebox{90}{\textbf{Kuwait}}} & \textbf{Overall Index} & \textbf{3.07} & \textbf{3.35} & -- \\
    \midrule
    
    \rowcolor{gray!10} & S1: Trust output & 2.69 (36) & 3.07 (41) & .134 \\
    & S2: Expert-level & 3.17 (36) & 3.29 (41) & .398 \\
    \rowcolor{gray!10} & S3: Predictability & 3.61 (36) & 3.56 (41) & .836 \\
    & S4: Uncertain & 2.81 (36) & 2.95 (41) & .690 \\
    \rowcolor{gray!10} & S5: Preference & 3.58 (36) & 3.39 (41) & .302 \\
    & S6: Overall trust & 3.33 (36) & 3.39 (41) & .716 \\
    \rowcolor{gray!10} \multirow{-7}{*}{\rotatebox{90}{\textbf{Jordan}}} & \textbf{Overall Index} & \textbf{3.20} & \textbf{3.28} & -- \\
    \bottomrule
  \end{tabular}
\end{table}

\begin{figure*}[t]
    \centering
    \includegraphics[width=0.9\textwidth]{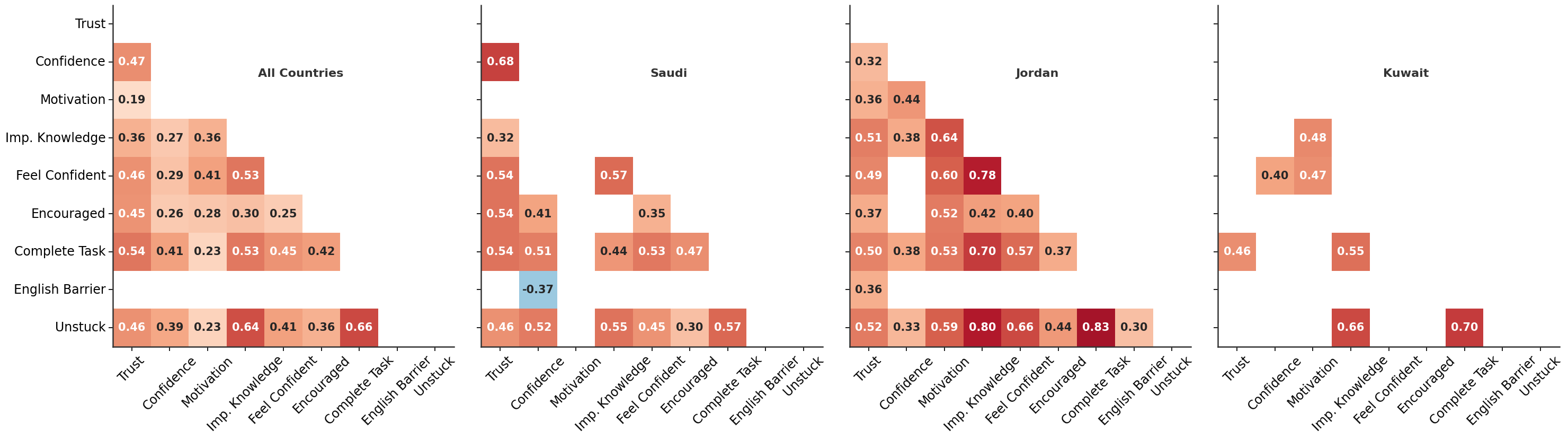}
    \caption{Correlation Matrix of Trust, Confidence, and Motivation Factors ($P-value < 0.01$). Empty cells indicate no significant correlations.}
    \label{fig:Correlation1}
\end{figure*}

Across the three Middle Eastern countries, Pearson correlation analysis reveals insights in Figure \ref{fig:Correlation1}. The highest correlation was found is between 'Getting Unstuck' and both 'Completing Tasks' (r=0.66) and 'Improved Knowledge' (r=0.64). This means that these tools help students solve problems, boosting both their productivity and learning. Trust plays a supporting role, correlating strongly with 'Completing Tasks' (r=0.54). This suggests that students trust AI tools because they help them finish their work, while the deeper pedagogical value comes from the tools' ability to provide solutions for programming challenges. In Saudi Arabia, an interesting finding is the significant negative correlation (-0.37) between 'English Barrier' and 'Confidence', which appeared only this country. This means that higher confidence is closely linked to a lower perceived language barrier. The strongest positive correlation in Saudi Arabia was between Trust and Confidence (r=0.68), potentially indicating that trust in AI may boost students' confidence. We also found strong positive correlations between Trust and 'Feeling More Confident' (r=0.54) as well as 'Being Encouraged' (r=0.54). Moreover, there was a correlation between 'Getting Unstuck' (r=0.55) and 'Completing Tasks' (r=0.57). In Jordan, we found very high correlations between 'Getting Unstuck' and 'Completing Tasks' (r=0.83), as well as between 'Getting Unstuck' and 'Improved Knowledge' (r=0.80). Additionally, there was a positive correlation between 'Improved Knowledge' and 'Motivation' (r=0.64). In Kuwait, there is a high correlation between 'Getting Unstuck' and 'Completing Tasks' (r=0.70), as well as between 'Getting Unstuck' and 'Improved Knowledge' (r=0.66).

\subsection{RQ3: Perceived Impact of AI on Programming Learning}

\subsubsection{Perceived Learning Outcomes}
Generally, students reported positive outcomes from AI usage across the six dimensions. Table \ref{tab:Outcomes} shows that the highest-rated outcome was statement O3 (Mean=4.03, SD=0.82), with 83.8\% of students agreeing or strongly agreeing. Statement O2 followed closely (Mean=3.88, SD=0.87), then statement O6 (Mean=3.77, SD=1.04). The language barrier item, statement O5, revealed mixed experiences; 44.4\% of students agreed they experienced such challenges, while 36.4\% disagreed. Interestingly, there is a relatively high standard deviation (1.32), suggesting considerable variability in language proficiency among students in each country.

\begin{table*}[htbp]
  \caption{Perceived Outcomes of AI Use ($N=198$)}
  \label{tab:Outcomes}
  \centering
  \small
  \renewcommand{\arraystretch}{1.3} 
  \setlength{\tabcolsep}{3pt} 
  
  
  \newcolumntype{Q}{>{\raggedright\arraybackslash\hsize=2.1\hsize}X}
  
  \newcolumntype{L}{>{\centering\arraybackslash\hsize=1.0\hsize}X}
  
  \newcolumntype{S}{>{\centering\arraybackslash\hsize=0.45\hsize}X}
  

  \begin{tabularx}{\textwidth}{ Q L L L L L S S }
    \toprule
    \textbf{Outcome Statement} & 
    \textbf{\shortstack{Strongly\\Disagree}} & 
    \textbf{Disagree} & 
    \textbf{Neutral} & 
    \textbf{Agree} & 
    \textbf{\shortstack{Strongly\\Agree}} & 
    \textbf{Mean} & 
    \textbf{\textit{SD}} \\
    \midrule
    
    \rowcolor{gray!10}
    O1: Helps complete tasks & 
    8 (4.0\%) & 11 (5.6\%) & 30 (15.2\%) & 121 (61.1\%) & 28 (14.1\%) & 
    3.76 & 0.91 \\
    
    O2: Helps when stuck & 
    7 (3.5\%) & 10 (5.1\%) & 16 (8.1\%) & 132 (66.7\%) & 33 (16.7\%) & 
    3.88 & 0.87 \\
    
    \rowcolor{gray!10}
    O3: Improves knowledge & 
    2 (1.0\%) & 11 (5.6\%) & 19 (9.6\%) & 113 (57.1\%) & 53 (26.8\%) & 
    4.03 & 0.82 \\
    
    O4: Increases confidence & 
    1 (0.5\%) & 20 (10.1\%) & 35 (17.7\%) & 102 (51.5\%) & 40 (20.2\%) & 
    3.81 & 0.89 \\
    
    \rowcolor{gray!10}
    O5: English language challenges & 
    33 (16.7\%) & 39 (19.7\%) & 38 (19.2\%) & 60 (30.3\%) & 28 (14.1\%) & 
    3.06 & 1.32 \\
    
    O6: Improved motivation & 
    12 (6.1\%) & 9 (4.5\%) & 35 (17.7\%) & 99 (50.0\%) & 43 (21.7\%) & 
    3.77 & 1.04 \\
    
    \bottomrule
    \multicolumn{8}{l}{\footnotesize \textit{Note:} \textit{SD} in the last column denotes Standard Deviation.}
  \end{tabularx}
\end{table*}

\subsubsection{Gender Differences in Outcomes by Country}

Table \ref{tab:outcome_gender_country} shows a gender gap across five of the six outcomes (all $p < 0.05$ except motivation) in Saudi Arabia, where male students regularly report 0.50–0.75 points higher outcomes than female students. It is evident that female students faced greater challenges regarding language barriers; for statement O5 ("When using AI, I face English language challenges"), they reported higher scores than males (3.72 vs. 3.12, $p = 0.033$). This aligns with the observed trust gap and suggests that segregated educational environments may create differential access to AI benefits. Furthermore, in statement O2, there is a 0.75-point gap between females and males, suggesting that female students may face a double disadvantage: struggling more with English and feeling less supported when they get stuck. Moreover, statement O1 ("Using AI helps me complete programming tasks") shows the largest gap between genders (0.85). Kuwait, another Gulf culture but with mixed-gender education, shows males trending only slightly higher; the only significant difference was in statement O6, where males reported higher motivation ($p = 0.030$). This suggests that while Saudi and Kuwaiti students share similar cultural backgrounds, a mixed-gender environment appears to equalize most outcomes. Supporting this, Jordanian male and female students representing a non-Gulf culture and mixed-gender education scored nearly identical (differences < 0.20) with no significant gender differences on any outcome (all $p > 0.05$). Jordan's results show the highest level of gender parity among the three countries, indicating that when cultural and environmental barriers are removed, gender differences in AI outcomes diminish.

\begin{table}[htbp]
  \centering
  \caption{Gender Differences in Outcomes by Country}
  \label{tab:outcome_gender_country}
  \footnotesize
  \renewcommand{\arraystretch}{1.3} 
  \setlength{\tabcolsep}{4pt} 
  
  \begin{tabular}{l l c c c}
    \toprule
    \textbf{Country} & 
    \textbf{Outcome} & 
    \textbf{\shortstack{Female\\Mean}} & 
    \textbf{\shortstack{Male\\Mean}} & 
    \textbf{\textit{p}} \\
    \midrule
    
    \rowcolor{gray!10}
      & O1: Complete tasks & 3.09 & 3.94 & .002 \\
      & O2: Helps when stuck & 3.40 & 4.15 & $<$ .001 \\
    \rowcolor{gray!10}
      & O3: Improves knowledge & 3.67 & 4.24 & .010 \\
      & O4: Increases confidence & 3.44 & 4.12 & $<$ .001 \\
    \rowcolor{gray!10}
      & O5: Language challenges & 3.72 & 3.12 & .033 \\
    \multirow{-6}{*}{\rotatebox{90}{\textbf{Saudi}}} 
      & O6: Improved motivation & 3.84 & 3.76 & .477 \\
    \midrule
    
    \rowcolor{gray!10}
      & O1: Complete tasks & 3.68 & 4.00 & .226 \\
      & O2: Helps when stuck & 3.84 & 4.05 & .083 \\
    \rowcolor{gray!10}
      & O3: Improves knowledge & 4.04 & 4.26 & .059 \\
      & O4: Increases confidence & 3.68 & 4.05 & .096 \\
    \rowcolor{gray!10}
      & O5: Language challenges & 2.76 & 2.63 & .712 \\
    \multirow{-6}{*}{\rotatebox{90}{\textbf{Kuwait}}} 
      & O6: Improved motivation & 3.44 & 3.84 & .030 \\
    \midrule
    
    \rowcolor{gray!10}
      & O1: Complete tasks & 3.94 & 4.07 & .675 \\
      & O2: Helps when stuck & 3.92 & 4.07 & .399 \\
    \rowcolor{gray!10}
      & O3: Improves knowledge & 4.00 & 4.12 & .562 \\
      & O4: Increases confidence & 3.75 & 3.95 & .628 \\
    \rowcolor{gray!10}
      & O5: Language challenges & 2.86 & 2.85 & .457 \\
    \multirow{-6}{*}{\rotatebox{90}{\textbf{Jordan}}} 
      & O6: Improved motivation & 3.75 & 3.88 & .527 \\
      
    \bottomrule
  \end{tabular}
\end{table}

\subsubsection{Qualitative Insights: Student Perceptions}
For this question, we used qualitative analysis. 82 out of 202 participants provided written responses to the open-ended question \textit{"In general, how do you feel about AI systems in programming?"}. Sentiment analysis revealed that most students had a positive outlook, with 62.2\% expressing positive feelings, 34.1\% neutral, and just 3.7\% negative. Interestingly, no significant differences were observed between genders (females: 62.5\% positive; males: 61.9\% positive) or between first-generation and continuing-generation students (first-generation: 60.9\% positive; continuing-generation: 62.7\% positive). Our thematic analysis identified five major themes:

\begin{itemize}

\item \textbf{Usefulness (n=68)}
In 57.1\% of the responses, students mostly focused on how useful AI is. Some examples include: \textit{"Excellent and helps me a lot, saving me a lot of time"}, \textit{"Good and explains in all programming languages"} and \textit{"Makes tasks very easy"}.

\item \textbf{Learning Support (n=23)}
In 19.3\% of the responses, students viewed AI as an educational tool and here are two examples for student responses: \textit{"Using it to explain code, concepts, and clarify outputs is acceptable and helps in developing programming skills"} and \textit{"It helps in understanding many things related to university subjects. It is a good reference and sometimes gives good ideas"}.

\item \textbf{Trust and Verification Concerns (n=13)} 
In 10.9\% of the responses, students showed and expressed healthy skepticism. For example, \textit{"Do not always trust it"}, \textit{"Good, but I do not trust it much"} and \textit{"It is good and helpful, but I have to review everything because it is not always accurate"}.

\item \textbf{Limitations and Risks (n=9)} 
In 7.6\% of the responses, students mentioned potential drawbacks. For instance: \textit{"Useful but overuse is dangerous"}, \textit{"It is helpful, but it does not always lead to real learning"} and \textit{"Using it to write the entire code will contain some errors... students will not be able to understand the code or improve later on"}.

\item \textbf{Time Saving Benefits (n=6)}
In 5.0\% of the responses, students appreciated how much time or effort could be saved, making tasks quicker and easier. Here are some responses: \textit{"Excellent when time is not available"} and \textit{"There are many benefits to using it, the most important of which is accomplishing tasks and saving effort and time"}. 

\end{itemize}

Students consistently emphasized that AI should supplement, not replace human judgment. As a student said: \textit{"I think it is a tool for assistance, nothing more and nothing less, and one should not rely on it too much."}

\subsubsection{Trust Rankings Based on Scenarios}
We provided students with two hypothetical scenarios to analyze how students would seek help in different contexts. The first scenario, \textit{"Code It Hackathon"} was focused on a relatively simple programming task, while the second scenario, \textit{"Digital Health Hackathon"}, was focused on more complex and high-stakes tasks. The goal of these differing tasks is to determine whether the perceived difficulty and potential consequences of each task influenced the students' preferences for seeking help. We want to explore the nature of each task that might affect the types of resources students would prefer to select (see Table \ref{tab:ranking_resources}). In the first scenario, students mostly trusted their own previously written code, followed by AI tools. For the second scenario, students again preferred their own previously written code, followed by expert developer code. This suggests that students build their trust based on task complexity. They prefer to use AI tools for simpler tasks but seek expert guidance for more complex, higher stakes tasks.

\begin{table}[htbp]
  \caption{Ranking of Coding Resources by Scenario}
  \label{tab:ranking_resources}
  \centering
  \footnotesize
  \renewcommand{\arraystretch}{1.3} 
  \setlength{\tabcolsep}{6pt} 
  
  \begin{tabular}{l c c}
    \toprule
    & \textbf{``Code It''} & \textbf{``Digital Health''} \\
    \textbf{Resource} & \textbf{Avg Rank} & \textbf{Avg Rank} \\
    \midrule
    
    \rowcolor{gray!10}
    Code I previously wrote & 2.25 & 2.62 \\
    
    AI tools (ChatGPT, Copilot) & 3.01 & 3.25 \\
    
    \rowcolor{gray!10}
    Teammate's code & 3.38 & 3.56 \\
    
    Expert developer code & 3.85 & 2.92 \\
    
    \rowcolor{gray!10}
    Code from internet & 3.93 & 4.11 \\
    
    Write from scratch & 4.58 & 4.54 \\
    
    \bottomrule
    \multicolumn{3}{l}{\footnotesize \textit{Note:} Lower numbers indicate better rankings (1 = Best, 6 = Worst).}
  \end{tabular}
\end{table}

\section{Discussion}

We distributed our survey to computing education departments at two universities in Saudi Arabia, one university in Kuwait and Jordan. Our study examined AI trust and usage patterns among 202 computer science students, and three major findings emerged: (1) 98\% of students across all three countries have used AI, reflecting very high adoption rates, with the primary usage being for learning rather than code generation; (2) moderate overall trust levels (Mean=3.11/5) with a statistically significant difference between genders in Saudi Arabia, but not in Kuwait or Jordan; and (3) students generally reported positive learning outcomes. Notably, Saudi students faced challenges with language barriers which limited some of the benefits they could gain. These results show how cultural, educational, and linguistic factors in the Middle East affect how students use AI tools. 

\paragraph{High Adoption Despite Cultural Differences}
Despite the segregated education environment in Saudi Arabia and the mixed-gender education environment in Kuwait and Jordan, where Jordan is a non-Gulf culture, both male and female students adopted AI at equally high rates. We found that students primarily use AI to understand the code (84.3\%) and learning new CS concepts (77.3\%) rather than writing new code (60.1\%). 

This trend may be driven by the linguistic context. Arabic-speaking students face an additional cognitive load when learning programming, which occurs mostly in English. Using AI to ``understand and explain code'' is very helpful, especially for non-native English speakers who need support with both language and technical concepts. This is consistent with Prather et al. \cite{prather_breaking_2024}’s findings on multilingual challenges when using AI.

However, small differences appeared in social factors. Saudi students were less likely than Jordanian students to believe that ``professionals use AI'', potentially due to Saudi Arabia’s more conservative professional culture. This may reflect adoption more slowly. 

The observed behaviors and the "healthy skepticism" towards AI among Middle Eastern students can be deeply understood through Hofstede’s Cultural Dimensions Theory \cite{hofstede2010cultures}. Arab nations traditionally score high on Uncertainty Avoidance, meaning individuals within these cultures often feel uncomfortable with ambiguity and untested situations. This cultural trait may provide a strong theoretical explanation for why Middle Eastern students, despite high adoption rates, maintain moderate and cautious trust levels compared to their Western peers who may exhibit more uncritical reliance. Furthermore, the higher Power Distance prevalent in these societies may influence how students view the authority of AI systems versus traditional human instructors. Framing our findings within this established framework illustrates that the varying levels of trust in AI are not merely geographical coincidences, but are deeply rooted cultural responses to uncertainty and authority in the face of new technology.

\subsection{Trust and Education Environment }

An interesting finding of this study is the fundamental gender gap in trust in Saudi Arabia. This special and unique pattern that appeared in Saudi Arabia needs careful interpretation within the country unique educational and social landscape. The academic environment in Saudi Arabia has two different environments for male and female students, which align with its cultural norms. However, this separation may inadvertently create variations in how students connect with other peers and the industry, or maybe female students may prefer working with their same gender peers rather than using AI and feel more comfortable. Male students in Saudi Arabia may have more exposure to technology and industry connections. Also, the technology sector in Saudi Arabia remains male dominated, which could potentially influence female students on their familiarity and trust in new tools like AI.

This is further influenced by linguistic and cultural factors. Female students reported higher language challenges (3.72 vs. 3.12). In addition, in conservative cultures, the academic expectations regarding academic excellence may influence female students' approach toward new technologies like AI tools, which potentially leading them to a more cautious approach to protect their academic performance. As a result, female students may encounter a combination of challenges: limitation of peer learning opportunities, weaker English preparation and limited technology industry exposure, which could influence their approach to adopting new technologies, potentially resulting in developing trust more slowly. 

The absence of gender differences in Kuwait and Jordan may indicate that sharing the educational environment leads to more equal exposure to new technologies and peer support for both genders. Male and female students likely benefit from this environment, by sharing the same classroom experiences, faculty interactions and even peer networks. This shared educational context appears to align their levels of trust and adoption of AI. Even in Jordan, which is a non-Gulf country, the fact that many women work in technology may create a unique condition where female students develop trust in AI similarly to male students.

To address these disparities and given that AI is everywhere, academic institutions in Saudi Arabia may consider enhancing targeted support for female students. This could involve exposing them more to industry professionals who successfully use AI and organizing workshops to build more technical confidence. Also, they can create weekly or monthly programs where more experienced female students share positive AI experiences. This could help bridge the gap in trust in order to gain more and equal benefits from AI technologies for all students.

\paragraph{Perceived Learning Outcomes }
Students across all three countries reported strong positive learning outcomes, particularly for knowledge improvement (Mean=4.03) and support when stuck (Mean=3.88). In the qualitative analysis, 62.2\% of students showed positive sentiment and reported using AI practically rather than blindly. They considered AI a tool that still requires human judgment, as seen in comments such as: ``A great help for the programmer, not a replacement,'' ``just for assistance,'' and ``I am the one who thinks and organizes the information.'' Based on these responses, students understand that AI tools are useful but have limits. Their focus on ``getting help, not replacing'' may reflect their academic integrity, self-improvement goals, or conservative cultural values.

\section{Threats to validity}
Our study has several limitations: 
the sample size is small, it is very difficult to generalize our findings to broader populations or the entire region. 
The responses that we obtained in this study show only a portion of the undergraduate and postgraduate students from four universities in Saudi Arabia, Kuwait and Jordan. Since our sample size is small, it is very difficult to generalize our findings to broader populations or the entire region. 

Our survey questions were based on previous studies. We gave the student the right to add more on \textit{“other”} field, so they can provide any answer that we do not expect. Since we translated our survey into Arabic, we had multiple sessions with different student backgrounds who speak Arabic and English. Our pilot study helped us to modify our questionnaire based on the suggestions of the participants.  To reduce fatigue, we tried to make the survey as short as possible.

\section{Conclusion}
This research examines AI trust and usage among computer science students in three Middle Eastern countries. Our study demonstrates that high adoption does not necessarily imply high trust, and that trust is influenced by various cultural and educational factors. We found that while 98\% of students use AI tools, they do so with "healthy skepticism."

Our findings indicate that the comparison between the three Middle Eastern students' perspectives of using AI is not all the same. The observed trust disparity found in Saudi Arabia's distinct academic environments compared to the shared educational system in Kuwait and Jordan, highlights the significant role of educational context in shaping student trust and confidence in using new technologies. This suggests that using AI in education should not follow a one approach, but instead use strategies that fit local academic systems.


\begin{acks}

\end{acks}

\bibliographystyle{ACM-Reference-Format}
\bibliography{references.bib}

@misc{sukiennik_evaluation_2025,
  title     = {An {Evaluation} of {Cultural} {Value} {Alignment} in {LLM}},
  url       = {http://arxiv.org/abs/2504.08863},
  doi       = {10.48550/arXiv.2504.08863},
  language  = {en},
  urldate   = {2025-11-18},
  publisher = {arXiv},
  author    = {Sukiennik, Nicholas and Gao, Chen and Xu, Fengli and Li, Yong},
  month     = apr,
  year      = {2025},
  note      = {arXiv:2504.08863 [cs]}
}

@article{hartley_artificial_2024,
  title      = {Artificial {Intelligence} {Supporting} {Independent} {Student} {Learning}: {An} {Evaluative} {Case} {Study} of {ChatGPT} and {Learning} to {Code}},
  volume     = {14},
  issn       = {2227-7102},
  shorttitle = {Artificial {Intelligence} {Supporting} {Independent} {Student} {Learning}},
  url        = {https://www.mdpi.com/2227-7102/14/2/120},
  doi        = {10.3390/educsci14020120},
  language   = {en},
  number     = {2},
  urldate    = {2025-11-18},
  journal    = {Education Sciences},
  author     = {Hartley, Kendall and Hayak, Merav and Ko, Un Hyeok},
  month      = jan,
  year       = {2024},
  pages      = {120}
}

@inproceedings{sarsa_automatic_2022,
  title     = {Automatic {Generation} of {Programming} {Exercises} and {Code} {Explanations} using {Large} {Language} {Models}},
  url       = {http://arxiv.org/abs/2206.11861},
  doi       = {10.1145/3501385.3543957},
  language  = {en},
  urldate   = {2025-11-18},
  booktitle = {Proceedings of the 2022 {ACM} {Conference} on {International} {Computing} {Education} {Research} - {Volume} 1},
  author    = {Sarsa, Sami and Denny, Paul and Hellas, Arto and Leinonen, Juho},
  month     = aug,
  year      = {2022},
  note      = {arXiv:2206.11861 [cs]},
  pages     = {27--43}
}

@misc{prather_breaking_2024,
  title      = {Breaking the {Programming} {Language} {Barrier}: {Multilingual} {Prompting} to {Empower} {Non}-{Native} {English} {Learners}},
  shorttitle = {Breaking the {Programming} {Language} {Barrier}},
  url        = {http://arxiv.org/abs/2412.12800},
  doi        = {10.48550/arXiv.2412.12800},
  language   = {en},
  urldate    = {2025-11-18},
  publisher  = {arXiv},
  author     = {Prather, James and Reeves, Brent N. and Denny, Paul and Leinonen, Juho and MacNeil, Stephen and Luxton-Reilly, Andrew and Orvalho, João and Alipour, Amin and Alfageeh, Ali and Amarouche, Thezyrie and Kimmel, Bailey and Wright, Jared and Blake, Musa and Barbre, Gweneth},
  month      = dec,
  year       = {2024},
  note       = {arXiv:2412.12800 [cs]}
}

@article{tayan_considerations_2024,
  title      = {Considerations for adapting higher education technology courses for {AI} large language models: {A} critical review of the impact of {ChatGPT}},
  volume     = {15},
  issn       = {26668270},
  shorttitle = {Considerations for adapting higher education technology courses for {AI} large language models},
  url        = {https://linkinghub.elsevier.com/retrieve/pii/S266682702300066X},
  doi        = {10.1016/j.mlwa.2023.100513},
  language   = {en},
  urldate    = {2025-11-18},
  journal    = {Machine Learning with Applications},
  author     = {Tayan, Omar and Hassan, Ali and Khankan, Khaled and Askool, Sanaa},
  month      = mar,
  year       = {2024},
  pages      = {100513}
}

@misc{amoozadeh_student-ai_2024,
  title      = {Student-{AI} {Interaction}: {A} {Case} {Study} of {CS1} students},
  shorttitle = {Student-{AI} {Interaction}},
  url        = {http://arxiv.org/abs/2407.00305},
  doi        = {10.48550/arXiv.2407.00305},
  language   = {en},
  urldate    = {2025-11-18},
  publisher  = {arXiv},
  author     = {Amoozadeh, Matin and Nam, Daye and Prol, Daniel and Alfageeh, Ali and Prather, James and Hilton, Michael and Ragavan, Sruti Srinivasa and Alipour, Mohammad Amin},
  month      = oct,
  year       = {2024},
  note       = {arXiv:2407.00305 [cs]}
}

@inproceedings{aljohani_evaluating_2025,
  address    = {Riyadh, Saudi Arabia},
  title      = {Evaluating {LLMs} for {Arabic} {Code} {Summarization}: {Challenges} and {Insights} from {GPT}-4},
  copyright  = {https://doi.org/10.15223/policy-029},
  isbn       = {979-8-3315-3969-6},
  shorttitle = {Evaluating {LLMs} for {Arabic} {Code} {Summarization}},
  url        = {https://ieeexplore.ieee.org/document/10908735/},
  doi        = {10.1109/CDMA61895.2025.00017},
  language   = {en},
  urldate    = {2025-11-18},
  booktitle  = {2025 8th {International} {Conference} on {Data} {Science} and {Machine} {Learning} {Applications} ({CDMA})},
  publisher  = {IEEE},
  author     = {Aljohani, Ahmed and Alharbi, Raed and Alkhaldi, Asma and Aljedaani, Wajdi},
  month      = feb,
  year       = {2025},
  pages      = {67--72}
}

@inproceedings{jacobs_evaluating_2024,
  title     = {Evaluating the {Application} of {Large} {Language} {Models} to {Generate} {Feedback} in {Programming} {Education}},
  url       = {http://arxiv.org/abs/2403.09744},
  doi       = {10.1109/EDUCON60312.2024.10578838},
  language  = {en},
  urldate   = {2025-11-18},
  booktitle = {2024 {IEEE} {Global} {Engineering} {Education} {Conference} ({EDUCON})},
  author    = {Jacobs, Sven and Jaschke, Steffen},
  month     = may,
  year      = {2024},
  note      = {arXiv:2403.09744 [cs]},
  pages     = {1--5}
}

@article{dickey_innovating_2024,
  title      = {Innovating {Computer} {Programming} {Pedagogy}: {The} {AI}-{Lab} {Framework} for {Generative} {AI} {Adoption}},
  volume     = {5},
  issn       = {2661-8907},
  shorttitle = {Innovating {Computer} {Programming} {Pedagogy}},
  url        = {http://arxiv.org/abs/2308.12258},
  doi        = {10.1007/s42979-024-03074-y},
  language   = {en},
  number     = {6},
  urldate    = {2025-11-18},
  journal    = {SN Computer Science},
  author     = {Dickey, Ethan and Bejarano, Andres and Garg, Chirayu},
  month      = jul,
  year       = {2024},
  note       = {arXiv:2308.12258 [cs]},
  pages      = {720}
}

@inproceedings{demidova_john_2024,
  address    = {Bangkok, Thailand},
  title      = {John vs. {Ahmed}: {Debate}-{Induced} {Bias} in {Multilingual} {LLMs}},
  shorttitle = {John vs. {Ahmed}},
  url        = {https://aclanthology.org/2024.arabicnlp-1.18},
  doi        = {10.18653/v1/2024.arabicnlp-1.18},
  language   = {en},
  urldate    = {2025-11-18},
  booktitle  = {Proceedings of {The} {Second} {Arabic} {Natural} {Language} {Processing} {Conference}},
  publisher  = {Association for Computational Linguistics},
  author     = {Demidova, Anastasiia and Atwany, Hanin and Rabih, Nour and Sha’ban, Sanad and Abdul-Mageed, Muhammad},
  year       = {2024},
  pages      = {193--209}
}

@misc{raihan_large_2024,
  title      = {Large {Language} {Models} in {Computer} {Science} {Education}: {A} {Systematic} {Literature} {Review}},
  shorttitle = {Large {Language} {Models} in {Computer} {Science} {Education}},
  url        = {http://arxiv.org/abs/2410.16349},
  doi        = {10.48550/arXiv.2410.16349},
  language   = {en},
  urldate    = {2025-11-18},
  publisher  = {arXiv},
  author     = {Raihan, Nishat and Siddiq, Mohammed Latif and Santos, Joanna C. S. and Zampieri, Marcos},
  month      = oct,
  year       = {2024},
  note       = {arXiv:2410.16349 [cs]}
}

@misc{sibaee_llms_2024,
  title     = {{LLMs} as {Compiler} for {Arabic} {Programming} {Language}},
  url       = {http://arxiv.org/abs/2403.16087},
  doi       = {10.48550/arXiv.2403.16087},
  language  = {en},
  urldate   = {2025-11-18},
  publisher = {arXiv},
  author    = {Sibaee, Serry and Najar, Omar and Ghouti, Lahouri and Koubaa, Anis},
  month     = mar,
  year      = {2024},
  note      = {arXiv:2403.16087 [cs]}
}

@article{arbab_students_2024,
  title      = {Student’s {Utilization} and {Assistance} of {AI} {Tools} in {Assessment} {Completion}: {Perceptions} and {Implications}},
  volume     = {7},
  copyright  = {http://creativecommons.org/licenses/by/4.0},
  issn       = {2576-2982, 2576-2974},
  shorttitle = {Student’s {Utilization} and {Assistance} of {AI} {Tools} in {Assessment} {Completion}},
  url        = {https://j.ideasspread.org/index.php/ilr/article/view/1256},
  doi        = {10.30560/ilr.v7n3p1},
  language   = {en},
  number     = {3},
  urldate    = {2025-11-18},
  journal    = {International Linguistics Research},
  author     = {Arbab, Atif Noor and Dhuhli, Badar Al and Krishnan, Yugesh and Crisostomo, Anna Sheila},
  month      = oct,
  year       = {2024},
  pages      = {p1}
}

@inproceedings{alpizar-chacon_students_2025,
  address   = {Nijmegen Netherlands},
  title     = {Student's {Use} of {Generative} {AI} as a {Support} {Tool} in an {Advanced} {Web} {Development} {Course}},
  isbn      = {979-8-4007-1567-9},
  url       = {https://dl.acm.org/doi/10.1145/3724363.3729106},
  doi       = {10.1145/3724363.3729106},
  language  = {en},
  urldate   = {2025-11-18},
  booktitle = {Proceedings of the 30th {ACM} {Conference} on {Innovation} and {Technology} in {Computer} {Science} {Education} {V}. 1},
  publisher = {ACM},
  author    = {Alpizar-Chacon, Isaac and Keuning, Hieke},
  month     = jun,
  year      = {2025},
  pages     = {312--318}
}

@article{pelaez-sanchez_impact_2024,
  title      = {The impact of large language models on higher education: exploring the connection between {AI} and {Education} 4.0},
  volume     = {9},
  issn       = {2504-284X},
  shorttitle = {The impact of large language models on higher education},
  url        = {https://www.frontiersin.org/articles/10.3389/feduc.2024.1392091/full},
  doi        = {10.3389/feduc.2024.1392091},
  language   = {en},
  urldate    = {2025-11-18},
  journal    = {Frontiers in Education},
  author     = {Peláez-Sánchez, Iris Cristina and Velarde-Camaqui, Davis and Glasserman-Morales, Leonardo David},
  month      = jun,
  year       = {2024},
  pages      = {1392091}
}

@article{jost_impact_2024,
  title    = {The {Impact} of {Large} {Language} {Models} on {Programming} {Education} and {Student} {Learning} {Outcomes}},
  volume   = {14},
  issn     = {2076-3417},
  url      = {https://www.mdpi.com/2076-3417/14/10/4115},
  doi      = {10.3390/app14104115},
  language = {en},
  number   = {10},
  urldate  = {2025-11-18},
  journal  = {Applied Sciences},
  author   = {Jošt, Gregor and Taneski, Viktor and Karakatič, Sašo},
  month    = may,
  year     = {2024},
  pages    = {4115}
}

@misc{prather_widening_2024,
  title      = {The {Widening} {Gap}: {The} {Benefits} and {Harms} of {Generative} {AI} for {Novice} {Programmers}},
  shorttitle = {The {Widening} {Gap}},
  url        = {http://arxiv.org/abs/2405.17739},
  doi        = {10.48550/arXiv.2405.17739},
  language   = {en},
  urldate    = {2025-11-18},
  publisher  = {arXiv},
  author     = {Prather, James and Reeves, Brent and Leinonen, Juho and MacNeil, Stephen and Randrianasolo, Arisoa S. and Becker, Brett and Kimmel, Bailey and Wright, Jared and Briggs, Ben},
  month      = may,
  year       = {2024},
  note       = {arXiv:2405.17739 [cs]}
}

@article{schiff_education_2022,
  title      = {Education for {AI}, not {AI} for {Education}: {The} {Role} of {Education} and {Ethics} in {National} {AI} {Policy} {Strategies}},
  volume     = {32},
  issn       = {1560-4292, 1560-4306},
  shorttitle = {Education for {AI}, not {AI} for {Education}},
  url        = {https://link.springer.com/10.1007/s40593-021-00270-2},
  doi        = {10.1007/s40593-021-00270-2},
  language   = {en},
  number     = {3},
  urldate    = {2025-11-18},
  journal    = {International Journal of Artificial Intelligence in Education},
  author     = {Schiff, Daniel},
  month      = sep,
  year       = {2022},
  pages      = {527--563}
}

@inproceedings{pirzado_evaluating_2025,
  title      = {Evaluating {Language} {Dependency} in {Large} {Language} {Models}: {A} {Study} on {Programming} {Queries} in {English} and {Spanish}},
  isbn       = {978-628-96613-1-6},
  shorttitle = {Evaluating {Language} {Dependency} in {Large} {Language} {Models}},
  url        = {https://laccei.org/LACCEI2025-Mexico/meta/FP458.html},
  doi        = {10.18687/LACCEI2025.1.1.458},
  language   = {en},
  urldate    = {2025-11-18},
  booktitle  = {Proceedings of the 23rd {LACCEI} {International} {Multi}-{Conference} for {Engineering}, {Education} and {Technology} ({LACCEI}): "{Engineering}, {Artificial} {Intelligence}, and {Sustainable} {Technologies} in service of society"},
  publisher  = {Latin American and Caribbean Consortium of Engineering Institutions},
  author     = {Pirzado, Farman Ali and Ahmed, Awais and Ibarra-Vázquez, Gerardo and Terashima-Marin, Hugo},
  year       = {2025}
}

@misc{phung_generative_2023,
  title      = {Generative {AI} for {Programming} {Education}: {Benchmarking} {ChatGPT}, {GPT}-4, and {Human} {Tutors}},
  shorttitle = {Generative {AI} for {Programming} {Education}},
  url        = {http://arxiv.org/abs/2306.17156},
  doi        = {10.48550/arXiv.2306.17156},
  language   = {en},
  urldate    = {2025-11-18},
  publisher  = {arXiv},
  author     = {Phung, Tung and Pădurean, Victor-Alexandru and Cambronero, José and Gulwani, Sumit and Kohn, Tobias and Majumdar, Rupak and Singla, Adish and Soares, Gustavo},
  month      = aug,
  year       = {2023},
  note       = {arXiv:2306.17156 [cs]}
}

@inproceedings{bassner_iris_2024,
  title      = {Iris: {An} {AI}-{Driven} {Virtual} {Tutor} {For} {Computer} {Science} {Education}},
  shorttitle = {Iris},
  url        = {http://arxiv.org/abs/2405.08008},
  doi        = {10.1145/3649217.3653543},
  language   = {en},
  urldate    = {2025-11-18},
  booktitle  = {Proceedings of the 2024 on {Innovation} and {Technology} in {Computer} {Science} {Education} {V}. 1},
  author     = {Bassner, Patrick and Frankford, Eduard and Krusche, Stephan},
  month      = jul,
  year       = {2024},
  note       = {arXiv:2405.08008 [cs]},
  pages      = {394--400}
}

@misc{amoozadeh_trust_2024,
  title      = {Trust in {Generative} {AI} among students: {An} {Exploratory} {Study}},
  shorttitle = {Trust in {Generative} {AI} among students},
  url        = {http://arxiv.org/abs/2310.04631},
  doi        = {10.48550/arXiv.2310.04631},
  language   = {en},
  urldate    = {2025-11-18},
  publisher  = {arXiv},
  author     = {Amoozadeh, Matin and Daniels, David and Nam, Daye and Kumar, Aayush and Chen, Stella and Hilton, Michael and Ragavan, Sruti Srinivasa and Alipour, Mohammad Amin},
  month      = feb,
  year       = {2024},
  note       = {arXiv:2310.04631 [cs]}
}

@article{pitts_understanding_nodate,
  title    = {Understanding {Human}-{AI} {Trust} in {Education}},
  language = {en},
  author   = {Pitts, Griffin and Motamedi, Sanaz}
}

@inproceedings{cambaz_use_2024,
  address    = {Portland OR USA},
  title      = {Use of {AI}-driven {Code} {Generation} {Models} in {Teaching} and {Learning} {Programming}: a {Systematic} {Literature} {Review}},
  isbn       = {979-8-4007-0423-9},
  shorttitle = {Use of {AI}-driven {Code} {Generation} {Models} in {Teaching} and {Learning} {Programming}},
  url        = {https://dl.acm.org/doi/10.1145/3626252.3630958},
  doi        = {10.1145/3626252.3630958},
  language   = {en},
  urldate    = {2025-11-18},
  booktitle  = {Proceedings of the 55th {ACM} {Technical} {Symposium} on {Computer} {Science} {Education} {V}. 1},
  publisher  = {ACM},
  author     = {Cambaz, Doga and Zhang, Xiaoling},
  month      = mar,
  year       = {2024},
  pages      = {172--178}
}

@article{garcia-mendez_review_2025,
  title    = {A review on the use of large language models as virtual tutors},
  volume   = {34},
  issn     = {0926-7220, 1573-1901},
  url      = {http://arxiv.org/abs/2405.11983},
  doi      = {10.1007/s11191-024-00530-2},
  language = {en},
  number   = {2},
  urldate  = {2025-11-18},
  journal  = {Science \& Education},
  author   = {García-Méndez, Silvia and Arriba-Pérez, Francisco de and Somoza-López, María del Carmen},
  month    = apr,
  year     = {2025},
  note     = {arXiv:2405.11983 [cs]},
  pages    = {877--892}
}

@article{chen_systematic_nodate,
  title    = {A {Systematic} {Review} on {Prompt} {Engineering} in {Large} {Language} {Models} for {K}-12 {STEM} {Education}},
  language = {en},
  author   = {Chen, I-Sheng and Wang, Danyang and Xu, Luyi and Cao, Chen and Fang, Xiao and Lin, Jionghao}
}

@inproceedings{hou_all_2025,
  address    = {Nijmegen Netherlands},
  title      = {'{All} {Roads} {Lead} to {ChatGPT}': {How} {Generative} {AI} is {Eroding} {Social} {Interactions} and {Student} {Learning} {Communities}},
  isbn       = {979-8-4007-1567-9},
  shorttitle = {'{All} {Roads} {Lead} to {ChatGPT}'},
  url        = {https://dl.acm.org/doi/10.1145/3724363.3729024},
  doi        = {10.1145/3724363.3729024},
  language   = {en},
  urldate    = {2025-11-18},
  booktitle  = {Proceedings of the 30th {ACM} {Conference} on {Innovation} and {Technology} in {Computer} {Science} {Education} {V}. 1},
  publisher  = {ACM},
  author     = {Hou, Irene and Man, Owen and Hamilton, Kate and Muthusekaran, Srishty and Johnykutty, Jeffin and Zadeh, Leili and MacNeil, Stephen},
  month      = jun,
  year       = {2025},
  pages      = {79--85}
}

@inproceedings{WEIRD,
  author    = {Linxen, Sebastian and Sturm, Christian and Br\"{u}hlmann, Florian and Cassau, Vincent and Opwis, Klaus and Reinecke, Katharina},
  title     = {How WEIRD is CHI?},
  year      = {2021},
  isbn      = {9781450380966},
  publisher = {Association for Computing Machinery},
  address   = {New York, NY, USA},
  url       = {https://doi.org/10.1145/3411764.3445488},
  doi       = {10.1145/3411764.3445488},
  booktitle = {Proceedings of the 2021 CHI Conference on Human Factors in Computing Systems},
  articleno = {143},
  numpages  = {14},
  location  = {Yokohama, Japan},
  series    = {CHI '21}
}

@inproceedings{korber2019,
  author    = {Moritz Körber},
  title     = {Theoretical considerations and development of a questionnaire to measure trust in automation},
  booktitle = {Proceedings of the 20th Congress of the International Ergonomics Association (IEA 2018)},
  pages     = {13--30},
  year      = {2019},
  publisher = {Springer}
}

@book{Corbin,
  title     = {Basics of Qualitative Research: Techniques and Procedures for Developing Grounded Theory},
  author    = {Corbin, J. and Strauss, A.},
  isbn      = {9781483315683},
  lccn      = {2014048844},
  url       = {https://books.google.com/books?id=hZ6kBQAAQBAJ},
  year      = {2014},
  publisher = {SAGE Publications}
}

@inproceedings{Haque,
  author    = {Haque, Summit and Hundhausen, Christopher},
  title     = {Generative AI Access, Usage, and Perceptions: An Empirical Comparison of Computing Students In The United States and Bangladesh},
  year      = {2025},
  isbn      = {9798400713408},
  publisher = {Association for Computing Machinery},
  address   = {New York, NY, USA},
  url       = {https://doi.org/10.1145/3702652.3744231},
  doi       = {10.1145/3702652.3744231},
  booktitle = {Proceedings of the 2025 ACM Conference on International Computing Education Research V.1},
  pages     = {109–124},
  numpages  = {16},
  location  = {
               },
  series    = {ICER '25}
}

@book{hofstede2010cultures,
  title     = {Cultures and organizations: Software of the mind, 3rd edition},
  author    = {Hofstede, Geert and Hofstede, Gert Jan and Minkov, Michael},
  year      = {2010},
  publisher = {McGraw-Hill}
}

\end{document}